\documentclass[11pt]{amsart}   

\usepackage{amsmath,amssymb,amsthm,psfrag}
\usepackage{epsfig,graphicx,verbatim}
\usepackage{color,psfrag}
\usepackage[bookmarks, bookmarksopen=true, bookmarksnumbered=true]{hyperref}
\numberwithin{equation}{section}

\newcounter{mycount}

\def\1{\mbox{1\hskip-.25em l}}
\newcommand{\beq}{\begin{equation}}
\newcommand{\eeq}{\end{equation}}

\begin{document}

\title{Spectral converters and luminescent solar concentrators} \maketitle
\smallskip
\author{Petra F. Scudo}{*},
\author{Luigi Abbondanza},
\author{Roberto Fusco}\\
\address{Eni S.p.A, \textit{Research Center for Non-Conventional Energies- Istituto ENI Donegani}}\\
\address{Via G.Fauser 4 - 28100 Novara (Italy)}\\
\email {*email for correspondence:}{petra.scudo@eni.it}\\
\bigskip
\section*{Abstract}
In this paper we present a comprehensive theoretical description of molecular spectral converters in the
specific context of Luminescent Solar Concentrators (LSCs). The theoretical model is an extension to a three-level system interacting with a solar radiation bath of the standard quantum theory of atomic radiative processes.
We derive the equilibrium equations of the conversion process and provide specific examples of application of this principle to the development of solar concentration devices.
\bigskip
\section{Introduction}

Luminescent solar concentrators (LSCs) have been first introduced in the late seventies as
one of the simplest methods of concentrating sunlight in thin polymeric slabs \cite{bat} and have been lately reconsidered in the light of the recent advances in material science and nano-optics as a cheap alternative to standard photovoltaic modules \cite{reisf,baldo,golds}.
The concentration process can be understood through the principles of classic geometric optics.
The polymeric slabs are doped with active molecules, which absorb a portion of the incident solar radiation and re-emit part of it at lower energies, thus performing a spectral down-shift. The radiation is isotropically scattered by the luminescent centers and is subjected to refraction upon reaching the interface between the slab and the air. Consequently, depending on the value of the material's refractive index, part of the radiation is trapped by total internal reflection and part of it is lost through the upper and lower surfaces. The trapped radiation is wave-guided and reaches the edges of the slab, where it can be converted into electricity by standard solar cells.
These systems do not require tracking and limit the surface of the high-cost materials used for the solar cells, thus offering a cheap alternative to standard photovoltaic modules.
However, to this day, a wide-scale application of these concentrators has been inhibited mainly by their low concentration/conversion efficiency, due to several loss mechanisms of which the main one is probably the low efficiency of the dyes' wave-guiding properties due to low absorption and self-absorption.
A comprehensive review on the current research status of LSCs is presented in \cite{sark}.
Different Authors considered the problem of modeling these devices, following either a thermodynamic approach \cite{therm} (based on Chandrasekhar's radiative transfer theory, \cite{chand}) or a computational ray-tracing approach \cite{rayt}.
Both ways represent a broad-scale, macroscopic theoretical model of LSCs.
In our paper, we focus instead on a microscopic model, aimed at characterizing the working principles of the device
at the molecular scale, considering the interactions between the doping dyes and the solar photons.
Indeed, the correct physical description underlying the trapping mechanism of a LSC relies on the principles of quantum mechanics which govern the molecular spectral conversion process. In doing this, we start off from the microscopic thermodynamic model developed by Ross \cite{ross} and later elaborated by Yablonovitch in his seminal works \cite{yablonovitch,y2}.
With respect to the latter models, we propose a modification of the molecular rate equations of the system
to correctly account for interaction processes involving multiple molecular levels.
We explicitly derive the equilibrium radiation density fields for the molecular system interacting with the solar photon bath and describe a method for calculating the parameters involved in the relaxation process coupled to the radiative interactions.

\section{Definitions and notations}

For a monocromatic light source $S$, we define the \textit{mean photon-flux density}
at frequency $\nu$, $\varphi(\nu)$, to be
\beq
\varphi(\nu)=\frac{I(\nu)}{h \, \nu},\eeq
where $h$ is the Planck constant and $I$ the light intensity.
The mean photon flux has units $photons/cm^2 \times s$ and may be expressed in terms
of the \textit{radiation energy density} per unit volume and unit bandwidth $\rho$ as
\beq
\varphi(\nu)=\frac{c \, \rho(\nu)}{h \, \nu},\eeq
where $c$ has to be substituted by $c/n$ for a material of refractive index $n$.
In our discussion we shall be focusing on the interactions between light and matter,
in the form of molecular or atomic transitions caused by a thermal radiation field impinging
on a set of molecules. The transitions between different molecular
energy levels due to electron excitations are related to photon absorption and emission.
A fundamental physical quantity of our model is represented
by the \textit{transition rate}, which expresses the probability density per unit time of a transition between atomic energy eigenstates
and is defined as
\beq
W= \varphi(\nu) \sigma(\nu),
\eeq
where $W$ is the transition rate and $\sigma(\nu)$ is the \textit{transition cross-section}, i.e.
the effective cross-sectional atomic area (in $cm^2)$.
As we shall see later, $\sigma$ is determined by the strength of the dipole oscillations which,
coupled to the incident photon momentum, give rise to the transition.
In a LSC the light beam inducing molecular transitions
comes from the sun and can be approximated by a thermal radiation distribution from
a black-body source of temperature $T\simeq 5800 K$.
The spectral energy density per unit volume and unit bandwidth is given by Planck's formula
for black-body radiation
\beq
\rho(\nu)= \frac{8 \, \pi \, h \, \nu^3}{c^3}\frac{1}{e^{h\, \nu/K \, T}-1}.\label{rhoterm}
\eeq
The same quantity can be expressed in terms of the wavelength $\lambda$ as
\beq
\rho(\lambda)= \frac{8 \, \pi \, h \, c}{\lambda^5}\frac{1}{e^{h\, c/\lambda K \, T}-1}.\label{rhoterm}
\eeq

\section{Interaction dynamics}

As mentioned earlier, whereas the coarse features of a LSC
can be treated with the principles of geometric optics, the correct dynamics of the
molecular interactions is described by the principles of quantum mechanics. The model developed here is of general validity and can be
applied to all cases in which fluorescent molecules act as individual spectral converters.
In the notation of second quantization, the electromagnetic field
is an ensemble of photons whose state is defined by
a momentum and polarization vector for each frequency $\nu$.
The field is specified by giving the number of photons in a given state; we set $n_{\mathbf{k}, \nu}$ to be the number of photons with momentum $\mathbf{k}$ and frequency $\nu$.

In a volume of space $V$ containing atoms or molecules the energy operator (total energy), or Hamiltonian $\mathcal{H}$, of the system
consists of three terms: the radiation field energy $\mathcal{H}_{em}$, the molecular energy $\mathcal{H}_{mol}$ and the interaction
energy $\mathcal{H}_{int}$.
The latter describes the intensity of the transition processes and is used to compute transition rates between different
energy levels.
We introduce the following notation borrowed from Yariv \cite{yariv}.
Let $\mathbf{E}_{\bf{k}\nu}$ be the electric field generated by photons of mode $\bf{k}\nu$ (polarization and frequency)
and $\bf{r}$ the position vector relative to a particle interacting with the field.
The interaction Hamiltonian can be written as
\begin{eqnarray}
\mathcal{H}_{int}= - e \mathbf{E}_{\bf{k}\nu} \cdot \bf{r}\nonumber\\
= i e \sqrt{\frac{h \nu_{\bf k}}{2 V \epsilon}} \left[ a^{\dagger}_{\bf{k}\nu}e^{-i \bf{k}\cdot \bf{r}} - a_{\bf{k}\nu}e^{i \bf{k}\cdot \bf{r}} \right] \mathbf{e}_{\bf{k}\nu} \cdot \bf{r} \label{Hint},
\end{eqnarray}
where $\epsilon$ is the electric permittivity, $a, a^{\dagger}$ are the photon annihilation and creation operator of a single radiation mode.
In our system, an incident radiation beam coming from the sun has a distribution of different frequencies and momenta, each associated to a specific molecular transition.
In general, when a photon of mode $\mathbf{k}\nu_1$ is absorbed, the molecular electrons are excited to a higher energy level: the
transition energy, which equals the electron's levels energy difference, is given by
\beq
E_1 - E_0= h \nu_1.
\eeq

We refer hereafter to a general chemical compound that can be schematized in terms of a three-level system:
a fundamental state and two excited states relative to absorption and emission. We consider the case in which these
levels may be treated as well-defined electronic eigenstates. Furthermore, we make the assumption- common in dealing with fluorescence- that spontaneous emission dominates over stimulated emission.

During absorption, the number of photons in the radiation field decreases by one unit (we consider only single-photon processes) while the molecular electronic state is shifted to an upper level. If $E_0, E_1$ are the lower and higher electronic energy eigenvalues respectively, the rate of absorption is given by \textit{Fermi's golden rule}
\beq
W_{abs}= h | \langle E_0, n_{\mathbf{k}\nu}|\mathcal{H}_{int}| E_1, n_{\mathbf{k}\nu}-1 \rangle|^2 \, \delta(E_1 - E_0 - h \nu_1),
\eeq
where $\delta$ denotes Dirac's delta function.
The number of photons in a given mode is related to the energy density by
\beq
n_{\mathbf{k}\nu}= \frac{\rho(\mathbf{k},\nu) V}{h\nu},
\eeq
and the average absorption rate is
\beq
\bar{W}_{abs} = \frac{2 \pi^2e^2\nu}{3 h^2 \epsilon} \, |\mu_{1,0}|^2 \, \rho(\nu),
\eeq
where $\mu_{1,0}$ is the dipole moment of the transition, and $\nu_1$ the associated frequency.
The same formula expresses also the rate of induced emission, where the electronic states return to a lower energy level by emitting one photon to the external field.
The coefficients of $W$ can be related to Einstein's coefficients of laser theory as follows.
According to Einstein \cite{einstein}, the rate of absorption
in the presence of thermal radiation is
\beq
W= B \rho(\nu),
\eeq
with $\rho(\nu)$ given by (\ref{rhoterm}).
The constant $B$ is expressed in terms of the above coefficients by
\beq
B= \frac{2 \pi^2 e^2}{3 \epsilon h^2} |\mu_{10}|^2,\label{B}
\eeq
and is termed \textit{absorption coefficient}. For an exhaustive discussion of Einstein's derivation and its implications on the quantum theory of light-matter interactions, see for example Dirac \cite{dirac}.
\section{Radiative processes}

The theoretical model developed hereafter applies in general to spectral conversion procedures in which absorption and emission involve different energies and different atomic configurations.
In most fluorescent species absorption and emission take place in two different parts of the molecule and involve different energy levels.
The first rigorous thermodynamic description of such spectral converters was formulated by Ross \cite{ross} and later extended by Yablonovitch \cite{yablonovitch,y2}.
However, their model was based on the assumption that both absorption and emission are fully reversible processes and can share chemical equilibrium with the radiation field. As we shall see later, a more correct description of a general molecular spectral converter, involves at least three, rather than only two, energy levels and therefore is characterized by a double-equilibrium condition between radiation-absorption and radiation-emission. These two processes, both reversible and thus entropy-preserving, are related to each other by an internal energy (or charge) transfer which requires the introduction of two different chemical potentials.
Associated to these processes there is an increase of entropy of the electronic system coupled with the photon bath:
the entropy acquired upon absorbing a photon of energy $h\nu_1$ is only partially rendered upon emission, as the emitted photon has an energy $h\nu_2$, with $\nu_2 < \nu_1$. This entropy difference can be more rigorously described referring to the change of molecular electronic states between absorption and emission by considering the transition from a pure electronic state to a statistical mixture of states due to a superposition of different nuclear wavefunctions. We can then apply the formula for the \textit{von Neumann entropy} $S$
of a quantum system \cite{nielsen} described by a density operator $\Sigma$
\begin{equation}
S(\Sigma)=-Tr{\Sigma \log(\Sigma)},
\end{equation}
where $\Sigma$ is the electronic density matrix corresponding to the emission excited states and the $\log$ function refers to its diagonal form.
Note that this analysis assumes that the state of the system before emission is described by a superposition of
nuclear and electronic wavefunctions of the form
\begin{equation}
\Psi=\sum_{l,m}c_{l,m}\Xi_l \otimes \Phi_{l,m},\label{sup}
\end{equation}
where $\Xi, \Phi$ denote the electronic, nuclear wavefunctions respectively. With the above notation, we have
\begin{equation}
\Sigma_{l,l'}= \sum_m \langle \Phi_{l,m}, \Psi \Psi^{\dagger} \Phi_{l',m} \rangle,
\end{equation}
where $\Psi \Psi^{\dagger}$ is the projector on the state (\ref{sup}).
The dynamics of the three processes (absorption, emission, thermalization) can be described by a set of differential
equations for the electronic occupations of the absorption and emission excited states respectively, here labelled as $N_1, N_2$
\begin{eqnarray}
\frac{d}{dt}N_1= B_1 \rho(\nu_1) N_0 - (B_1 \rho(\nu_1)+ A_1) N_1 - q N_1\\
\frac{d}{dt}N_2= B_2 \rho(\nu_2) N_0 - (B_2 \rho(\nu_2)+ A_2) N_2 + q N_1 \label{system1},\nonumber
\end{eqnarray}
where $B$ is given by (\ref{B}), $\frac{B}{A}=\frac{\lambda^3}{8 \pi h}$, and $q$ is the energy transfer intensity, specified below.

\begin{figure}[h!]
\centering
\includegraphics[width=8cm, height=5cm]{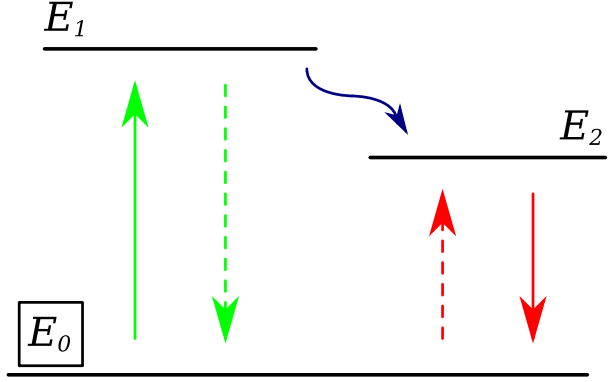}
\caption{Sketch of the process dynamics: absorption (between levels $E_0$ and $E_1$), spontaneous emission (between levels $E_2$ and $E_0$) and internal energy transfer (between levels $E_1$ and $E_2$). Processes occurring with low probability are represented by dashed lines and the energy levels refer to the electronic states exclusively.}
\end{figure}

$A$ is Einstein's coefficient of spontaneous emission, introduced in order to account for thermal equilibrium and is simply given by the inverse of the spontaneous life-time of the atomic level.
The system (\ref{system1}) can be solved by considering that, as thermodynamic equilibrium is reached,
the population fractions referred to the ground state are the relative Boltzmann distributions
\begin{eqnarray}
\frac{N_1}{N_0}= \exp[-(h \nu_1 - \mu_1)/\kappa T_m]\label{eqstates}\\
\frac{N_2}{N_0}= \exp[-(h \nu_2 - \mu_2)/\kappa T_m],\nonumber
\end{eqnarray}
where $T_m$ is referred to the molecular temperature after the interaction and $\mu_1, \mu_2$ are the equilibrium chemical potentials relative to $\nu_1, \nu_2$ respectively.

The set of equations can be solved for the energy densities yielding
\begin{eqnarray}
\rho(\nu_1)=\frac{8 \pi h \nu_1^3/c^3 + q/B_1}{e^{(h \nu_1 - \mu_1)/\kappa T_m}-1}\label{densities}\\
\rho(\nu_2)=\frac{8 \pi h \nu_2^3/c^3 - q/B_2 e^{(h \nu_1 - \mu_1 -h \nu_2 + \mu_2)/\kappa T_m}}
{e^{(h \nu_2 - \mu_2)/\kappa T_m}-1}.\nonumber
\end{eqnarray}

In the above formulas, we have introduced a frequency-dependent \textit{chemical potential} of the molecular system in equilibrium with the radiation field \cite{wurfel}.
With regard to the latter quantity (borrowed from the theory of lasers), it is important to observe that from a statistical point of view the molecular system is described by a Grand Canonical ensemble in thermal and chemical equilibrium with the photon bath of the radiation field. In this case, the standard concept of particle exchange is substituted by the more general concept of energy quanta exchange. Indeed, the chemical potential of a system constitutes a measure of its ability to exchange particles with a bath at assigned values of the other thermodynamic parameters.

 From (\ref{densities}) one can note that the presence of the dissipative term $q$ modifies the standard energy distributions by increasing the density of the incident radiation field, while decreasing the emitted one. Thus, the term $q$ denotes the rate of energy transfer from radiative to thermal modes due to interaction between light and particles.

This term can be related to the quantum-mechanical description of molecular dynamics as follows.
Our molecular system can be decomposed in three constituents: the electronic states involved with absorption, those involved with emission, and the remaining nuclear degrees of freedom related to equilibrium rotational and vibrational motions. The above electronic states are differently localized in the compound; emission states are generally related to a metallic ion.

The total molecular wave-function can be computed within the frame of the so called \textit{Born-Oppenheimer approximation}. The underlying idea of this approximation method consists in computing separately the electronic and the nuclear wavefunctions, the separation being justified by their different kinetic energies.
In the first step, the electronic wavefunctions are found by solving Schr\"{o}dinger's equation for fixed values of the nuclear coordinates
\beq
\mathcal{H}_e(r,R) \Xi_l (r,R) = E^e_l \Xi_l (r,R),
\eeq
where $E^e_l=E^e_l(R)$ are the electronic eigenvalues, dependent on the nuclear coordinates $R$. By varying adiabatically the value of $R$ one can determine the potential energy surface $E^e_l(R)$.
In the second step, the nuclear energies and wavefunctions are derived in correspondence of a given electronic eigenvalue by introducing in the Schr\"{o}dinger equation the kinetic term which was initially neglected
\beq
[T_n + E^e_l(R)] \Phi_{l,m}(R)= E^n_m \Phi_{l,m}(R),
\eeq
where $T_n$ contains partial derivatives with respect to $R$, which is now a variable and no longer a parameter.

 Thus, the state of the electrons depends parametrically on the nuclei positions, whereas the one of the nuclei is labeled by the electronic eigenvalues.
On the basis of these considerations, we can write the total approximate state of our molecule as a factorized wavefunction
\beq
\Xi_l (r,R) \otimes \Phi_{l,m}(R),\label{zeroorder}
\eeq
where
$r, R$ denote the electronic and nuclear coordinates respectively, $\Xi$ the electronic component and $\Phi$ the nuclear one.

Upon absorbing a photon of frequency $\nu$, the electronic energy levels of the molecule undergo a transition from a state $\Xi_{l_0}$ to a state $\Xi_{l_1}$, with $h \nu= E^e_{l_1}-E^e_{l_0}$. In parallel, interaction with the incident photons causes a change in the nuclear momentum with a relative increase of the kinetic energy, from a value $E_{m_1}$ to a value $E_{m_2}$. This transition corresponds to a vibrational motion of the nuclei away from their original equilibrium positions and a rearrangement of the global molecular coordinates. In turn, nuclear motion causes a perturbation of the electronic wavefunctions which determines a change in charge distribution. The latter may be described as an electron transfer which involves a (possibly) non-radiative transition from the original level $\Xi_{l_1}$, acquired after absorption, to an intermediate state $\Xi_{l_1}+\delta \Xi_{l_1}$, before finally reaching the emission state $\Xi_{l_2}$.
 The transition rate of this internal energy transfer can be estimated from the principles of perturbation theory applied to the action of the nuclear kinetic energy operator on the unperturbed wavefunctions (\ref{zeroorder})
\begin{equation}
q=|\Psi_i, T_n \Psi_f|^2 \delta(E_2-E_1),
\end{equation}
where $\Psi_i=\Xi_{l_1}(r,R) \otimes \Phi_{l_1,m_1}(R)$ is the total molecular state after absorption, $\Psi_f=\Xi_{l_2}(r,R) \otimes \Phi_{l_2,m_2}(R)$ the one before emission, and the delta function $\delta(E_2-E_1)$ refers to the total energy balance of the transfer including both the electronic and the nuclear shifts, with $E_{1,2}=E_{m_{1,2}}+E^e_{l_{1,2}}$.
The order of magnitude of q can be computed using quantum chemical methods, which allow to determine the overlap between the two molecular wavefunctions involved in the process.

 After reaching the state $\Xi_{l_2}$, the electrons return to their original energy level by emitting a photon of lower frequency $\nu'$, such that $h \nu'=E^e_{l_2}-E^e_{l_0}$.

\section{Incident and emitted radiations}

The incident and emitted energy densities are represented by (\ref{densities}),
with chemical potentials
\beq
\mu_i = E_i \left( 1- \frac{T_0}{T_1}\right),
\eeq
where $T_0, T_1$ are the ambient temperature and the incident field black-body temperature respectively.
In a real setting, the incident radiation causes thermal energy exchanges between the particles and the photon field.
After absorption, the system transfers energy to a lower electronic level and thus loses kinetic energy
in terms of heat by raising its temperature from $T_0$ to a higher value $T_m$. (Note that this temperature has been already introduced above as the correct equilibrium temperature of the molecules).
This process corresponds to thermalization of the molecular system due to non radiative
internal energy transfer (the rate of which is given by the term $q$ above).
$T_m$ can be estimated thanks to
the \textit{equipartition theorem} of statistical mechanics, which states
that in a system at equilibrium, the temperature is equally distributed between the
different forms of internal energy.
It follows that the molecular temperature $T_m$ can be computed by the internal energy average
using the relation
\beq
\bar{E}_{rv} \propto \kappa T_m,
\eeq
where $\kappa$ is Boltzmann's constant and $E_{rv}$ is the molecular roto-vibrational energy.
The latter quantity can also be calculated by quantum chemical simulations, as well as experimentally measured.
The relative numerical values and figures are included in a separate file.

\section{Concentration and efficiency}

From (\ref{densities}), we see that the energy density of the emitted field at the emission frequency is higher than the one of the incident field at the absorption frequency.
On the basis of the above treatment, we could estimate an ideal spectral concentration factor directly related to the conversion process,
defined by the relation
\beq
C_M= \frac{\rho_2}{\rho_1},
\eeq
which, neglecting stimulated emission, yields
\beq
C_M= \frac{8\pi h \nu_2^3/c^3 - q/B_2 e ^{(h \nu_1 - \mu_1 -h \nu_2 + \mu_2)/\kappa T_m}}{8\pi h \nu_1^3/c^3 + q/B_1} e^{\mu_2+\mu_1-h \nu_2-h \nu_1/\kappa T_m}.
\eeq

Ideal efficiencies of LSCs are however limited by
the combination of four loss mechanisms: absorption efficiency ($\eta_a$), quantum efficiency of fluorescence ($\eta_f$), self-absorption coefficient ($\theta_q$) and total internal reflection ($\eta_t$).

To give a concrete example, let us consider a slab of refractive index $n=1.5$ doped with a compound having $\eta_a=0.05$, $\eta_f=0.7$ and $\theta_q=0$.
Defining the geometric concentration factor $G$ as
the ratio between the upper, absorbing surface and the edge surface, we obtain the total flux gain as
\beq
G_{\Phi}= G \eta_a \eta_f \eta_t (1- \theta_q).
\eeq

 Above we report a plot of the incident solar spectrum as a function of wavelength and of the emitted fluorescence spectrum modulated by the molecular blackbody radiation which propagates toward the edges of the slab.

\begin{figure}
\centering
\includegraphics{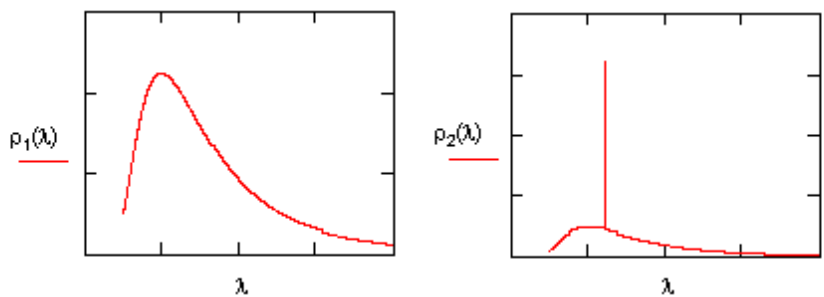}
\caption{Energy density vs wavelength of incoming (left) and wave-guided (right) radiations.}
\end{figure}

\section{Conclusions}

In this work we presented a microscopic model for the interaction processes taking place
in a LSC. These processes involve multiple molecular levels, as they are usually coupled to non-radiative, besides than radiative, energy transfer mechanisms. We derive the rate equations for the molecular system in equilibrium with the solar photon bath and highlight a way of determining the parameters involved in the non-radiative processes, which are a subject of a further investigation currently in progress. Furthermore, we provide examples of the type of spectral distributions involved.
In order to fully understand the efficiency of LSCs it appears fundamental to handle not only the macroscopic, ray-tracing aspect of the device, but also the molecular-scale efficiencies involved. In these respects, it would be of great importance to be able to predict on the basis of molecular properties the entity of thermal relaxation and fluorescence efficiency.

\subsection*{Acknowledgements}
P.F.S. would like to thank L.C. Andreani for helpful discussions and for useful comments on the manuscript.

\end{document}